\documentclass[journal]{IEEEtran}

\usepackage{cite}
\usepackage[aboveskip=0pt]{caption}
\usepackage{pifont}
\usepackage{booktabs}
\usepackage{amsmath}
\usepackage{float}
\usepackage{graphicx}
\usepackage{multirow}
\usepackage{algorithm}
\usepackage{algpseudocode}
\usepackage[colorlinks=true, allcolors=blue]{hyperref}

\usepackage[table]{xcolor}

\title{Classification based deep learning models for lung cancer and disease using medical images}

\author{Ahmad~Chaddad*, Jihao Peng, Yihang~Wu
\IEEEcompsocitemizethanks{\IEEEcompsocthanksitem This work did not involve human subjects or animals in its research.\\
This research was funded by the National Natural Science Foundation of China \#82260360, the Guilin Innovation Platform and Talent Program \#20222C264164, and the Guangxi Science and Technology Base and Talent Project (\#2022AC18004 and \#2022AC21040).\\
A. Chaddad, Y. Wu and J. Peng are with the Laboratory for Artificial Intelligence of Personalized Medicine, School of Artificial Intelligence, Guilin University of Electronic Technology, Guilin, China.\\
A. Chaddad is with The Laboratory for Imagery, Vision and Artificial Intelligence, Ecole de Technologie Superieure, Montreal, Canada.
\\All co-authors contributed equally. *Corresponding author: Ahmad Chaddad. \\
Email: ahmad8chaddad@gmail.com, ahmadchaddad@guet.edu.cn}}

\begin{document}
\maketitle
\begin{abstract}
\textcolor{black}{The use of deep learning (DL) in medical image analysis has significantly improved the ability to predict lung cancer. In this study, we introduce a novel deep convolutional neural network (CNN) model, named ResNet+, which is based on the established ResNet framework. This model is specifically designed to improve the prediction of lung cancer and diseases using the images. To address the challenge of missing feature information that occurs during the downsampling process in CNNs, we integrate the ResNet-D module, a variant designed to enhance feature extraction capabilities by modifying the downsampling layers, into the traditional ResNet model. Furthermore, a convolutional attention module was incorporated into the bottleneck layers to enhance model generalization by allowing the network to focus on relevant regions of the input images. We evaluated the proposed model using five public datasets, comprising lung cancer (LC2500 $n$=3183, IQ-OTH/NCCD $n$=1336, and LCC $n$=25000 images) and lung disease (ChestXray $n$=5856, and COVIDx-CT $n$=425024 images). To address class imbalance, we used data augmentation techniques to artificially increase the representation of underrepresented classes in the training dataset. The experimental results show that ResNet+ model demonstrated remarkable accuracy/F1, reaching 98.14/98.14\% on the LC25000 dataset and 99.25/99.13\% on the IQ-OTH/NCCD dataset. Furthermore, the ResNet+ model saved computational cost compared to the original ResNet series in predicting lung cancer images. The proposed model outperformed the baseline models on publicly available datasets, achieving better performance metrics. Our codes are publicly available at \url{https://github.com/AIPMLab/Graduation-2024/tree/main/Peng}.} 
\end{abstract}

\begin{IEEEkeywords}
Deep learning, ResNet, lung cancer, classification
\end{IEEEkeywords}

\section{Introduction}\label{sec:introduction}

Lung cancer is the most common and deadly cancer worldwide, with an estimated 2 million newly diagnosed cases and 1.8 million deaths \cite{background}. The incidence rate of lung cancer is highest in developing countries, where smoking is the most common. In addition, lung cancer is one of the leading causes of death in China. Since 2010, the incidence rate, mortality, and disease burden of lung cancer have increased \cite{backgroundCHN}. The improvement in survival rate can be attributed to early detection and improvements in treatment methods, including targeted therapy and immunotherapy \cite{10234092}. However, current diagnostic methods are invasive, time-consuming, and expensive and may not always provide an accurate diagnosis. It is important to develop a rapid and adaptable approach for the initial screening of patients to improve the diagnosis of lung cancer.

DL is a machine learning method based on artificial neural networks that can automatically extract features and perform accurate pattern recognition by learning large amounts of data. In lung cancer, DL can be applied to extract key features from CT scans, X-rays and other image data, and assist physicians in the early detection and diagnosis of lung cancer \cite{s4}. DL can be trained on a large number of lung imaging data to learn features of lung cancer, such as lumps and nodules \cite{intro3}. By distinguishing these features from normal tissue, models can accurately segment lung cancer lesions, helping physicians detect early \cite{10364730, 10521883}. It can also be combined with other clinical data, such as patient medical history and biomarkers, to build a multimodal prediction model \cite{intro4, 10336413}. For example, with the accumulation of increasingly more medical image data, the convolutional neural network (CNN) model can continuously learn from the data and continuously improve its performance \cite{intro6}. However, CNN-based lung cancer prediction faces several challenges. The ratio between normal and lung cancer samples is often unbalanced. The quality of images in a dataset can vary significantly, leading to noise or artifacts, which affects the performance of the CNN. Specifically, it can cause the model to predict the majority classes during training and to poorly predict minority classes. As CNNs are viewed as 'black box' models, it is a challenge for clinicians to understand the reasoning behind model predictions \cite{challage1}. This impacts the generalizability of the model. A CNN trained on one dataset may not generalize well to other unseen similar datasets, a difference that may be due to variations in imaging equipment \cite{challage2}.

In this paper, we propose new CNN models based on the residual neural network (ResNet), including 1) the improvement scheme proposed by ResNet-D \cite{ResNet_D} and 2) the convolutional attention mechanism \cite{CBAM}. The main contributions of this paper can be summarized as follows.

\begin{itemize}
    \item We involve the attention model to the ResNet50 and ResNet101 models for predicting the lung cancer images.
    \item We evaluate the impact of attention model with CNNs using public medical images.
    \item \textcolor{black}{We combine different modalities (e.g., CT with pathology) to validate our model for predicting lung cancer.}
\end{itemize}

The paper is structured as follows. First, Section \ref{sec:Related work} presents related work. In Section \ref{sec:METHODOLOGY}, we provide our proposed model. Next, we present the experimental results in Section \ref{sec:Experiment}. Finally, we discuss and conclude the paper in Section \ref{conc}.

\begin{figure*}
    \centering
    \includegraphics[width=0.93 \linewidth]{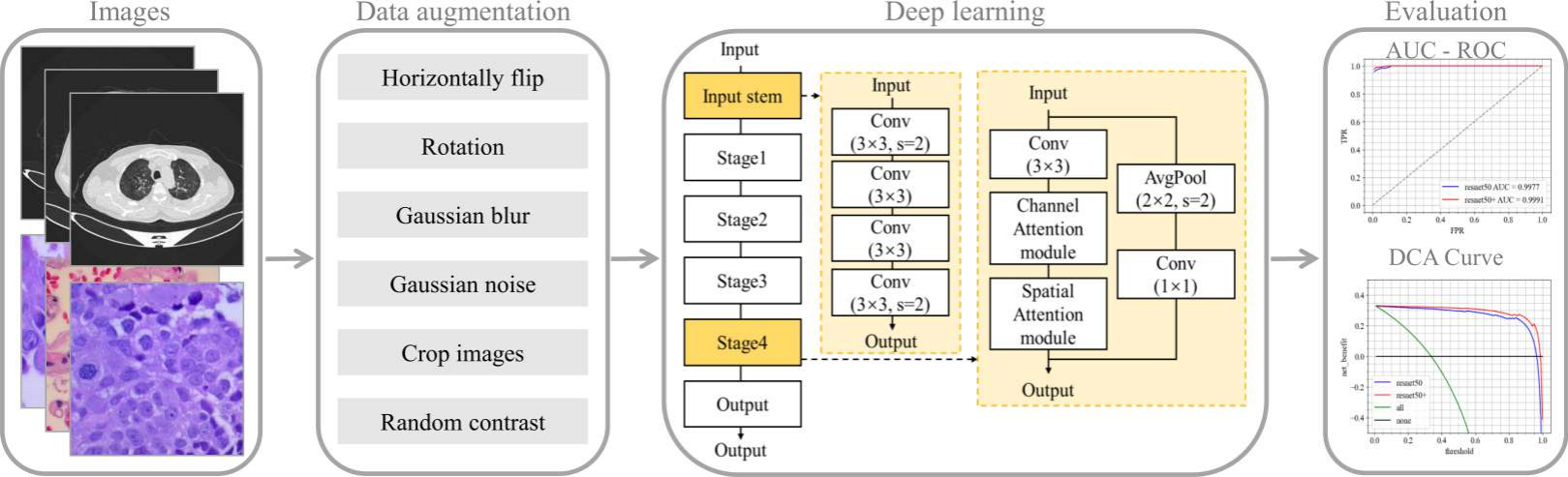}
    \caption{Diagram illustrating the workflow for classifying lung cancer images. It covers image acquisition, data augmentation, the classification model, and performance evaluation metrics. }
    \label{fig:pipline}
\end{figure*}

\section{Related Work}\label{sec:Related work}

Recently, CNN architectures have shown significant advantages in the classification and early detection of lung cancer. In \cite{related-1}, lung cancer diagnosis was achieved by analysis of CT and histopathological images, comparing six different CNN models, with accuracy reaching 97.86\%. \textcolor{black}{However, their method showed fluctuations in validation accuracy during training, suggesting it may not be robust for real-world applications.} Another study conducted a comparative analysis using seven pre-trained CNN models, highlighting ResNet 101 as the most accurate with 98.67\% \cite{related-10}.

Combining machine learning (ML) and DL models can mitigate the challenges posed by small sample sizes, enhancing the interpretability and flexibility of the model. A hybrid network to classify lung histopathological images achieved 99. 3\% accuracy by integrating discrete wavelet transform coefficients and deep features from AlexNet \cite{related-4}. An automatic nodule detection method in CT images, based on an improved AlexNet architecture and the Support Vector Machine (SVM) algorithm, achieved 97.64\% accuracy \cite{related-6}. \textcolor{black}{However, it introduces extra time to train the SVM classifiers.} Using DenseNet as the baseline, a study applied two feature selection methods to optimize the extraction of features from DenseNet201, achieving an average accuracy of 95\% with ML classifiers \cite{related-3}. \textcolor{black}{Yet, their study uses only small lung cancer datasets raises concerns about potential overfitting, which may limit the generalizability of their method.} Another study introduced a modified and optimized CNN architecture for automatic identification of lung cancer in CT images, demonstrating higher performance compared to alternative strategies \cite{related-7}. \textcolor{black}{However, their study lacks generalization to larger datasets, and training efficiency is insufficient due to high computational costs for hyperparameter optimization.} In \cite{related-8}, a novel deep CNN framework outperforms Inception V3, Xception, and ResNet50 models with accuracy of 92\%. The use of an ensemble method for the detection of lung nodules significantly improved the robustness of the model, achieving an overall accuracy of 95\% by integrating the output from multiple CNNs \cite{related-2}. Using the pre-trained VGG19 model and three CNN blocks for the feature extraction and classification stages, their network performance reached an accuracy of 96.48\% \cite{related-5}. \textcolor{black}{However, it needs to fine-tune $\sim 24.62$M parameters, which is time-consuming.}

Exploring multimodal CNN models improves the diversity and accuracy of early lung cancer screening. Changes in the composition and concentration of volatile organic compounds in exhaled breath positively impact early detection of lung cancer. A study developed a breath analysis system that uses a gas sensor array and DL algorithm, achieving 97.8\% accuracy in classifying healthy controls and patients with lung cancer using comprehensive clinical datasets \cite{related-9}. \textcolor{black}{Yet, a lack of validation on public available medical datasets limits its applicability.} In addition, multimodal sensor systems and CNNs have been explored for the diagnosis of lung cancer from patient time series CT scans, improving the assessment of the burden of cardiopulmonary disease, and uncovering under-recognized pathologies \cite{intro7}. Given these studies on lung cancer, further research is important to improve performance, especially when dealing with limited data sets. Attention mechanisms could potentially identify relevant hidden features of lung cancer, leveraging these features to enhance CNN classifier performance.

\textcolor{black}{Unlike previous studies, we introduce an attention model to the ResNet series to effectively improve its feature extraction ability. Furthermore, we employ data augmentation as a regularization to avoid the risk of overfitting.}

\section{METHODOLOGY}\label{sec:METHODOLOGY}
Figure \ref{fig:pipline} illustrates the flow chart of our model. We collect data sets from open-source public libraries. We perform data augmentation on datasets (i.e., IQ-OTH/NCCD) to avoid imbalance classes. The data sets collected are used to evaluate our CNN models (ResNet+) compared to the baseline ResNet50 and ResNet101 for the purpose of multiclass classification.

\subsection{Datasets}
\begin{figure}
\centering
\includegraphics[width=0.9 \linewidth]{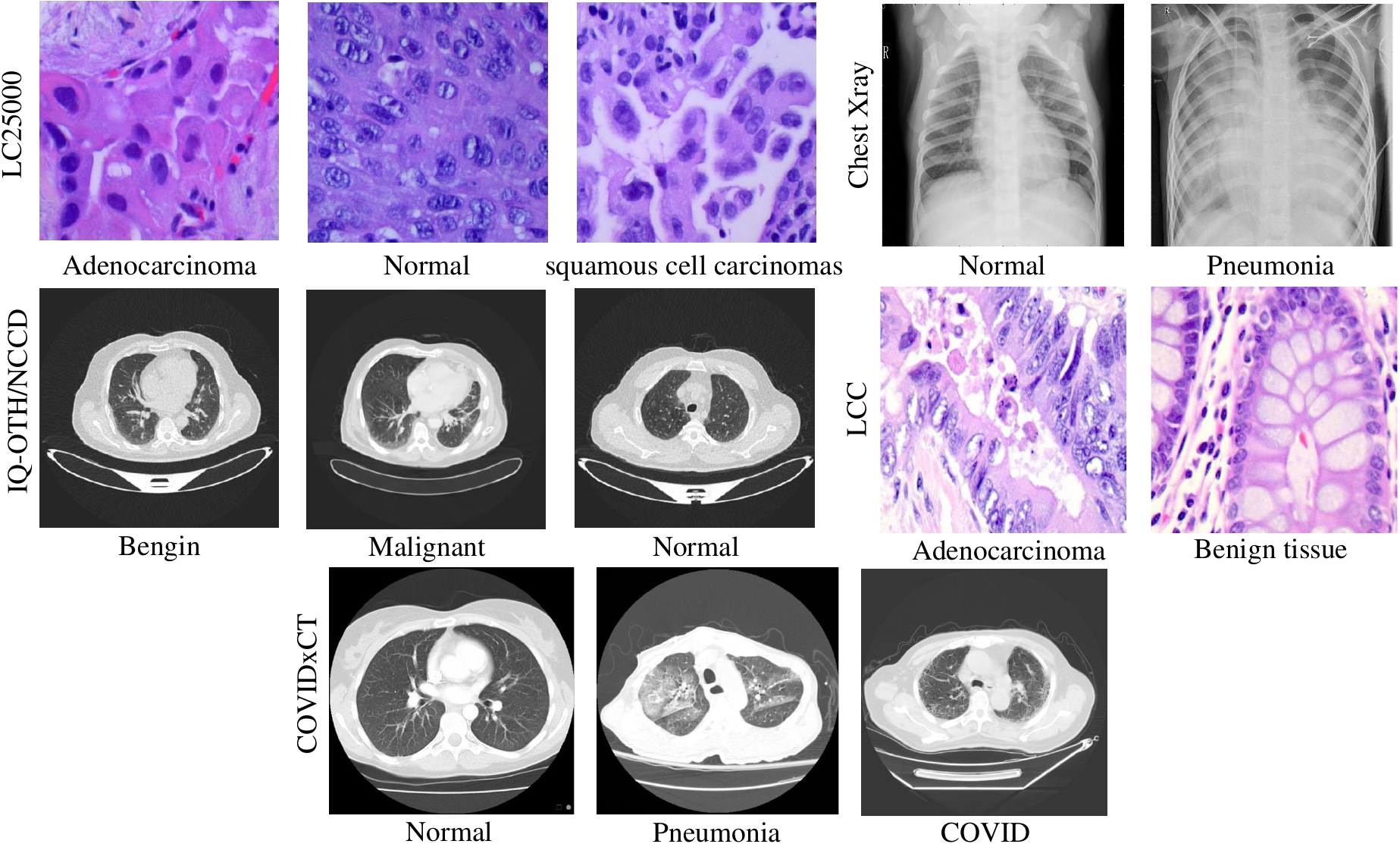}
\caption{Example images derived from LC25000, IQ-OTHNCDD, LCC, ChestXray and COVIDxCT datasets.}
\label{Img}
\end{figure}

We used histopathological images of lung cancer from the LC25000 dataset \cite{lc25000} and CT scans from the Iraq Cancer Teaching Hospital / National Center for Cancer Diseases (IQ-OTH/NCCD). Furthermore, we also considered COVID-related datasets for validation. Figure \ref{Img} illustrates a sample image from these datasets. 

\noindent\textbf{LC25000} consists of 15000 lung histopathological images (adenocarcinomas (Aca), lung squamous cell carcinomas (Scc) and benign lung tissues (Benign)), each sized $768 \times 768$ pixels in JPEG format. From this dataset, we randomly selected 3183 images for training, validation, and testing, distributed as follows: 2400 for training, 300 for validation, and 483 for testing.

\noindent\textbf{IQ-OTH/NCCD} consists of CT scans from patients diagnosed with various stages of lung cancer and healthy subjects, totaling 1097 labeled images \cite{alyasriy_hamdalla_2021}. \textcolor{black}{These CT scan slices are derived from 110 patients and grouped into three classes: normal (n=416), benign (n=120), and malignant (n=561).} Specifically, there are 40 malignant cases, 15 benign cases, and 55 normal cases. To address class imbalance in the IQ-OTH/NCCD dataset, we employed data augmentation techniques: 1) Horizontal Flip, 2) Affine Transformation, 3) Gaussian Blur, 4) Additive Gaussian Noise, 5) Cropping, 6) Linear Contrast adjustment, and 7) randomly applying transformations to process the images. Through data augmentation techniques, we used 1336 images for training, validation, and testing, with a distribution of 1070 for training, 133 for validation, and 133 for testing.

\noindent\textcolor{black}{\noindent\textbf{LCC} Lung and colon cancer (LCC) contains 25000 histopathological images with 5 classes, namely lung benign tissue (LBT), lung adenocarcinoma (LUAD), lung squamous cell carcinoma (LUSC), colon adenocarcinoma (COAD), colon benign tissue (CBT) \cite{borkowski2019lung}. The train set has 17500 images (LBT: 3500; LUAD: 3500; LUSC: 3500; COAD: 3500 CBT: 3500), the validation set has 2500 images (LBT: 500; LUAD: 500; LUSC: 500; COAD: 500 CBT: 500), and the test set contains 5000 images (LBT: 1000; LUAD: 1000; LUSC: 1000; COAD: 1000 CBT: 1000).}

\noindent\textcolor{black}{\noindent\textbf{ChestXray.} This dataset includes 5863 chest X-ray images (JPEG) organized into train, test, and val sets, with two classes, namely Pneumonia (train: n=3875; val: n=8; test: n=390) and Normal (train: n=1341; val: n=8; test: n=234) \cite{kermany2018identifying}. These data were collected from pediatric patients aged 1-5 years at the Guangzhou Women and Children Medical Center, and the images were examined and diagnosed by two physicians, with a third expert reviewing the evaluation set.}

\noindent\textbf{COVIDx-CT.} To assess the impact of the proposed model on non-cancer lung diseases, we also used COVIDx-CT, derived from CT imaging data collected by the China National Center for Bioinformation (CNCB) \cite{COVID}. This data set consists of chest CT examination data from various hospital cohorts in China, including images of novel coronavirus pneumonia, ordinary pneumonia, and normal lungs. \textcolor{black}{It consists of a total of 5312 patients, with 4249 for training (Normal: n=321; Pneumonia: n=592; COVID-19: n=3336), 560 for validation (Normal: n=164; Pneumonia: n=202; COVID-19: n=194), and 503 for testing (Normal: n=164; Pneumonia: n=138; COVID-19: n=201).}

\textcolor{black}{Table \ref{tab:data} reports the class distributions of images used for the training, validation (Val), and testing of CNN models using the five datasets. As illustrated, IQ-OTH/NCCD shows class imbalance compared to LC25000.}


\begin{table}
    \centering    
    \caption{{Dataset image distribution.}}
    \label{tab:data}
    \begin{tabular}{c|c|ccc|c}
    \hline
    
Name & Split & Aca & Scc & Benign & Total\\ \hline
                                  & Train& 800 & 800  & 800 & 2400\\
                                LC25000&Val  & 100 & 100  & 100 & 300\\
                                &Test & 161 & 161  &161  & 483\\ 
                                \cmidrule(l{3pt}r{3pt}){2-2}\cmidrule(l{3pt}r{3pt}){3-6}
& & Benign & Malignant & Normal & \\
\cmidrule(l{3pt}r{3pt}){2-2}\cmidrule(l{3pt}r{3pt}){3-6}
                                 &Train & 287 & 449 & 334 & 1070\\
                                 IQ-OTH/NCCD& Val &36 & 56 & 41 & 133\\
                                &Test & 36 & 56 & 41 & 133\\
                                \cmidrule(l{3pt}r{3pt}){2-2}\cmidrule(l{3pt}r{3pt}){3-6}
 & & Normal & Pneumonia & Covid & \\
 \cmidrule(l{3pt}r{3pt}){2-2}\cmidrule(l{3pt}r{3pt}){3-6}
                                 &Train & 35996 & 26970 & 294552 & 357518\\
                                COVIDx CT&Val & 17570 & 8008 & 8147 & 33725\\
                                &Test & 17922 & 7965 & 7894 & 33781\\ \cmidrule(l{3pt}r{3pt}){1-2}\cmidrule(l{3pt}r{3pt}){3-6}
    \end{tabular}
\setlength{\tabcolsep}{6.35pt}
\begin{tabular}{c|c|ccccc|c}
   & & LBT&LUAD&LUSC&COAD&CBT&\\
   \cmidrule(l{3pt}r{3pt}){2-2}\cmidrule(l{3pt}r{3pt}){3-8}
    \multirow{3}{*}{LCC}&Train &3500 &3500 &3500 &3500 &3500&17500 \\
& Val &500 &500 &500 &500 &500&2500 \\
    & Test &1000 &1000 &1000 &1000 &1000&5000 \\
    \cmidrule(l{3pt}r{3pt}){1-2}\cmidrule(l{3pt}r{3pt}){3-8}
\end{tabular}
\setlength{\tabcolsep}{12.2pt}
\begin{tabular}{c|c|cc|c}
     &&Pneumonia&Normal&\\
     \cmidrule(l{3pt}r{3pt}){2-2}\cmidrule(l{3pt}r{3pt}){3-5}
    \multirow{3}{*}{ChestXray}&Train &3875 &1341 &5216 \\
    &Val &8 &8 &16  \\
    &Test &390 &234 &624  \\
    \hline
\end{tabular}

\end{table}

\subsection{Proposed model}

Our proposed model builds upon the ResNet architecture tailored for classification tasks. We have integrated a Convolutional Block Attention Module (CBAM) into the bottleneck layer and made modifications to the network's downsampling structure. This enhanced variant of the residual neural network is denoted as ResNet+ (e.g., ResNet50+ or ResNet101+). In the following, we outline the details of our improved CNN architecture.

\begin{figure}
    \centering
    \includegraphics[width=0.9 \linewidth]{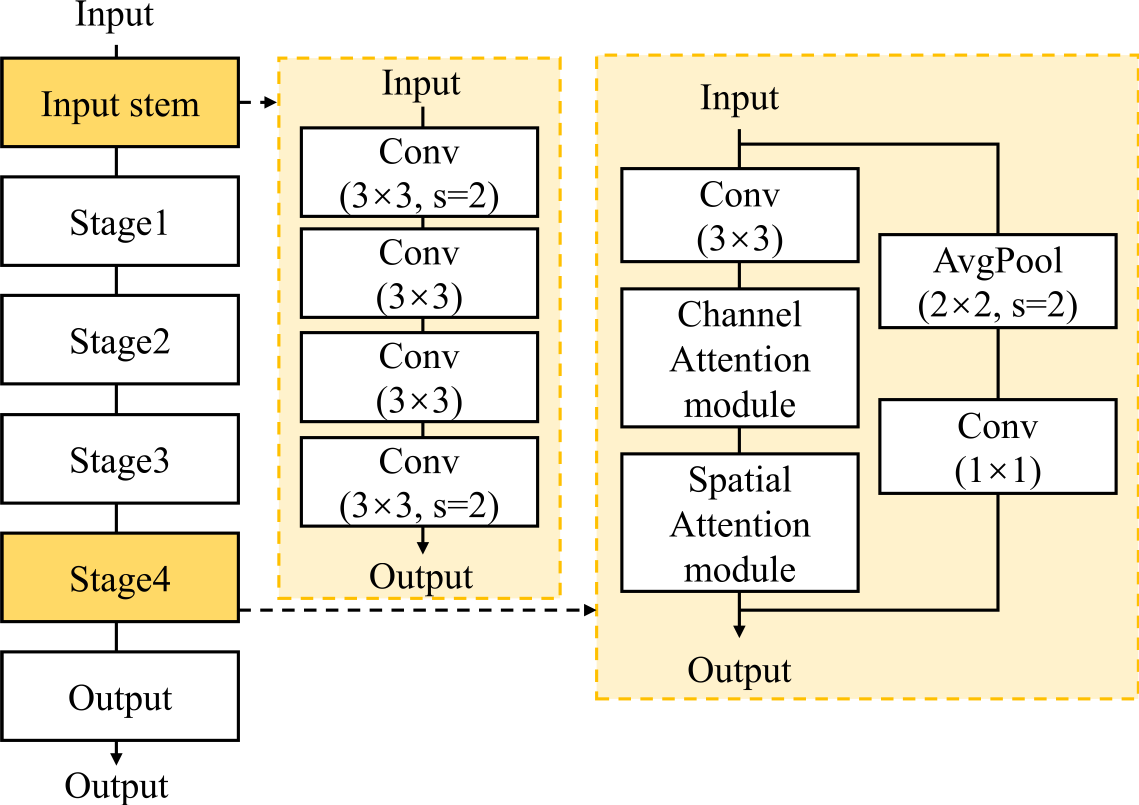}
    \caption{Structure of the classifier model includes stages with bottlenecks (e.g., ResNet50 with stages [3, 4, 6, 3]). The proposed model incorporates two modifications: ResNet-D and CBAM. \textcolor{black}{ResNet-D modifies the input stem and shortcut connections to enhance feature extraction efficiency. CBAM integrates channel attention and spatial attention modules to selectively focus on important features in the channel and spatial dimensions.}}
    \label{fig:network}
\end{figure}

\subsubsection{Convolutional block attention model}
The CBAM module has been integrated into the ResNet bottleneck layer to enhance the network's ability to learn dominant features. In CNNs, this module improves the network's comprehension of image features by combining channel attention and spatial attention mechanisms \cite{CBAM}. The CBAM module is designed to improve network performance and efficiency by dynamically adjusting the channel and spatial responses of the feature map. Specifically, the channel attention module learns the importance of each channel adaptively using two operations: squeeze and excitation. It weights the contribution of each channel in the feature map based on the task requirements \cite{SEnet}. This can be represented as follows:

\begin{equation}
\label{align:1}
F_{c}=\sigma(MLP(avgpool(M))+MLP(maxpool(M)))
\end{equation}

Where $\sigma$ denotes the sigmoid function, $MLP$ is a multilayer perceptron that consists of two fully connected layers, $M$ is a convoluted feature map. After the max-pooling and average-pooling operations, $M$ is entered into $MLP$, and finally added to the sigmoid. The pseudocode of the channel attention module is presented in the Algorithm \ref{tab:ChannelAttention}.

The purpose of spatial attention is to identify important areas in an image \cite{CBAM}. It counts the important information of the spatial position by pooling on each channel dimension, and learns the weights of each spatial position through a convolutional layer and a sigmoid function. Finally, weights are applied to each spatial location on the feature map to produce features with enhanced spatial importance. It can be expressed as follows.

\begin{equation}
\label{align:2}
 F_{s}=\sigma(Conv^{7\times7}(avgpool(F_{c}),maxpool(F_{c})))
\end{equation}

where $\sigma$ denotes the sigmoid function, $Conv^{7\times7}$ represents the convolution kernel of size $7 \times 7$, and $F_{s}$ is the output of the channel attention module. The pseudocode of the spatial attention module is presented in Algorithm \ref{tab:SpatialAttention}.

\subsubsection{ResNet-D}
We employ the ResNet-D architecture to enhance the preservation of the input feature map information. ResNet-D builds on the original ResNet architecture with three key improvements as proposed by \cite{ResNet_D}:
\begin{itemize}
 \item The first two convolutions in the bottleneck structure are swapped: a $1\times1$ convolution with stride 2 is followed by a $3\times3$ convolution with stride 1.

 \item The $7\times7$ convolution kernel between the input layer and the bottleneck layer is replaced by continuous $3\times3$ convolutions. This modification, inspired by changes in Inception-v2 \cite{inception}, aims to reduce computational costs.

 \item The shortcut downsampling module is adjusted: a $2\times2$ average pooling layer (stride 2) precedes the $1\times1$ convolutional layer. Subsequently, the stride of the $1\times1$ convolution is set to 1, preserving all input information with minimal computational impact.

 \item Finally, the stride-2 $3\times3$ maxpool layer before entering the first bottleneck layer is replaced by a $3\times3$ convolution.
\end{itemize}
These enhancements collectively improve the efficiency and effectiveness of ResNet-D for various classification tasks.

\subsubsection{Model process}
 Figure \ref{fig:network} illustrates the proposed model. Initially, images undergo data augmentation, cropping, and normalization, resulting in a size of $224\times224$. These preprocessed images are then fed into the stem layer, which transforms the image size from $(3,224,224)$ to a feature map of $(64,56,56)$. These feature maps are subsequently input into a series of four bottleneck layers that incorporate modifications from ResNet-D and integrate CBAM. Upon entering each bottleneck layer, the feature map undergoes convolution. Each feature map, sized $H \times W \times C$, then passes through the channel attention module. This module applies average and max pooling to the feature map, producing pooled outputs that are fed into fully connected layers to learn channel weights represented by $1 \times 1 \times C$, as detailed in Equation \ref{align:1}. These weights are multiplied by the corresponding post-activation channels to produce the output of the channel attention module.

Next, the spatial attention module processes the feature maps by performing average and max pooling across the channel dimension $C$, resulting in spatial weights of $H \times W \times 2$. These weights are combined to create a single $H \times W \times 1$ map, which is then multiplied element-wise with the input feature map, as described in Equation \ref{align:2}. The resulting feature map is combined with the downsampled shortcut connection, passes through the CBAM module, and is sent to the bottleneck layer. The detailed process of handling feature maps within each bottleneck layer is summarized in Algorithm \ref{tab:bottleneck}.

After passing through the four bottleneck layers, the feature map undergoes a random dropout and is finally fed into the fully connected layer, which maps it to the final classification output.

\begin{algorithm}\footnotesize
\caption{Channel Attention.}
\label{tab:ChannelAttention}
\begin{algorithmic}
\Procedure{ChannelAttention}{c, ratio}
\State $fc \gets$ nn.Sequential(nn.Conv2d(c, c // ratio, 1, bias=False), nn.ReLU(), nn.Conv2d(c // ratio, c, 1, bias=False))
\State $avg\_out \gets$ avg\_pool(x)
\State $max\_out \gets$ max\_pool(x)
\State $out \gets fc(avg\_out) + fc(max\_out)$
\State \textbf{return} sigmoid(out)
\EndProcedure
\end{algorithmic}
\end{algorithm}

\begin{algorithm}\footnotesize
\caption{Spatial Attention.}
\label{tab:SpatialAttention}
\begin{algorithmic}
\Procedure{SpatialAttention}{S}
\State $conv1 \gets$ nn.Conv2d(2, 1, S, padding=S//2, bias=False)
\State $avg\_out \gets$ mean(x)
\State $max\_out \gets$ max(x)
\State $x \gets$ torch.cat([avg\_out, max\_out], dim=1)
\State $x \gets$ conv1(x)
\State \textbf{return} sigmoid(x)
\EndProcedure
\end{algorithmic}
\end{algorithm}

\begin{algorithm}\footnotesize
\caption{BottleNeck Block.}
\label{tab:bottleneck}
\begin{algorithmic}
\Procedure{BottleNeckBlock}{c, c\_out, st}
\State $down\_sample \gets$ nn.Sequential(nn.AvgPool2d(2, 2), nn.Conv2d(c, c\_out, 1, 1), nn.BatchNorm2d(c\_out), nn.ReLU(c\_out))
\State $shortcut \gets$ down\_sample(x)
\State $x \gets$ conv1(x)
\State $x \gets$ conv2(x)
\State $x \gets$ conv3(x)
\State $x \gets$ ChannelAttention(x) * x
\State $x \gets$ SpatialAttention(x) * x
\State $output \gets$ relu(shortcut + x)
\State \textbf{return} output
\EndProcedure
\end{algorithmic}
\end{algorithm}

\subsection{Training strategy}
We used exponential moving averages (EMA), data augmentation, and cosine learning rate decay as training strategies \cite{morales2024exponential}. EMA is a weighted averaging regularization technique, which is an averaging method that gives higher weights to recent data to learn the flat optimal solution in deep neural network optimization to improve generalization ability. The decay of the cosine learning rate helps the model converge faster by dynamically adjusting the learning rate while avoiding falling into a local minimum, thus improving the generalization ability of the model \cite{cosine}.

\subsection{Performance Metrics}
The metrics used to assess deep CNN models for classification tasks include accuracy (ACC), precision (PRE), recall (REC), F1\_score, and confusion matrix. The detailed equations for the evaluation metrics are presented below:

\begin{equation}
Accuracy = \frac{TP + TN}{TP + TN + FP + FN}
\end{equation}

\begin{equation}
Precision = \frac{TP}{TP + FP}
\end{equation}

\begin{equation}
 Recall = \frac{TP}{TP + FN}
\end{equation}

\begin{equation}
 F1-Score = \frac{2 \times TP}{2 \times FP + FN}
\end{equation}

Where TP (True Positive), TN (True Negative), FP (False Positive), and FN (False Negative) denote the number of correct positives, correct negatives, incorrect positives, and incorrect negatives. 

Furthermore, receiver operating characteristic (ROC) curves with values of the area under the ROC curve (AUC) and decision curve analysis (DCA) \cite{fitzgerald2015decision} were used to evaluate the performance of the proposed model. Specifically, we calculate one-vs-all for ROC curves and DCA using scikit-learn libraries. 


\begin{table} \scriptsize
\caption{{Summary of the performance metrics (\%) for the classifications using the test sets.}}
 \setlength{\tabcolsep}{7pt}
 \renewcommand{\arraystretch}{1}
\begin{tabular}{c|c|ccccc}
\hline
                            dataset&Model& ACC & PRE & REC & F1&AUC\\ \hline
                            
\multirow{4}{*}{LC25000}&ResNet50+          & 98.14   & 98.14    & 98.14 & 98.14&0.998   \\
                            &   ResNet101+      & 97.52   & 97.53    & 97.52 & 97.52&0.996     \\
                            &   ResNet50        & 96.69   & 96.71    & 96.69 & 96.69&0.998   \\
                            &   ResNet101       & 96.07   & 96.40    & 96.07 & 96.06&0.995     \\ \cmidrule(l{3pt}r{3pt}){2-2}\cmidrule(l{3pt}r{3pt}){3-7}

 \multirow{4}{*}{IQ-OTHNCDD}&ResNet50+          & 99.25   & 99.21    & 99.07 & 99.13&0.999   \\
                            &     ResNet101+    & 98.50    & 98.42    & 98.18 & 98.25&0.999   \\
                            &     ResNet50      & 96.24    & 96.22   & 95.82 & 95.90&0.997   \\
                            &  ResNet101       & 97.74   & 98.02    & 97.45 & 97.71&0.996  \\ 
\cmidrule(l{3pt}r{3pt}){2-2}\cmidrule(l{3pt}r{3pt}){3-7}
\multirow{4}{*}{LCC} & ResNet50+     & 99.99   & 81.85    & 84.58 & 79.85&0.987  \\
   & ResNet101+    & 99.44    & 83.72    & 85.87 & 82.16&0.976   \\
   &   ResNet50  & 97.94    &  82.43  & 85.03 & 81.32&0.968   \\
   &  ResNet101   & 95.82   & 81.76    & 82.95 & 78.98&0.949   \\
\cmidrule(l{3pt}r{3pt}){2-2}\cmidrule(l{3pt}r{3pt}){3-7}
    \multirow{4}{*}{ChestXray}& ResNet50+ & 87.98 & 89.58 & 85.00 & 86.49 & 0.960 \\
& ResNet101+ & 83.01 & 81.79 & 82.82 & 82.19 & 0.895 \\
& ResNet50 & 84.94 & 88.25 & 80.60 & 82.47 & 0.931 \\
& ResNet101 & 79.97 & 85.08 & 73.97 & 75.56 & 0.868 \\
\cmidrule(l{3pt}r{3pt}){2-2}\cmidrule(l{3pt}r{3pt}){3-7}
             
 \multirow{4}{*}{COVIDxCT} & ResNet50+    & 78.45   & 81.85    & 84.58 & 79.85&0.973  \\
   &  ResNet101+   & 81.21    & 83.72    & 85.87 & 82.16&0.979   \\
    & ResNet50   & 80.58    &  82.43  & 85.03 & 81.32&0.969   \\
   &  ResNet101  & 77.55   & 81.76    & 82.95 & 78.98&0.961   \\
    \hline
                   
\end{tabular}
\label{tab:result}

+ sign after ResNet indicates the use of the ResNet-D architecture and the addition of CBAM.
\end{table}

\section{Results}\label{sec:Experiment}

\subsection{Experimental environment}\label{SS:implementation}
The experiments were performed on a system running the Windows 11 operating system with 128GB RAM, with an RTX 2060 GPU. CNN models were implemented using Python 3.9 with the PyTorch framework. \textcolor{black}{For training, the cross-entropy loss function was used for the classification task. The model was trained with a batch size of 16. The initial learning rate was set to $0.01$ and dynamically adjusted during the training process using a cosine annealing learning rate scheduler (maximum cycle length 40 epochs and a minimum learning rate of $1 \times 10^{-6}$). We used a stochastic gradient descent (SGD) optimizer with momentum of $0.9$. To enhance training stability, an EMA with a decay rate of 0.995 is applied to the model parameters. The model is trained for 200 epochs, and the validation model with the highest validation accuracy is selected for testing.}

\subsection{Model performance}

\begin{figure*}[!ht]
\centering
\includegraphics[width=0.98\linewidth]{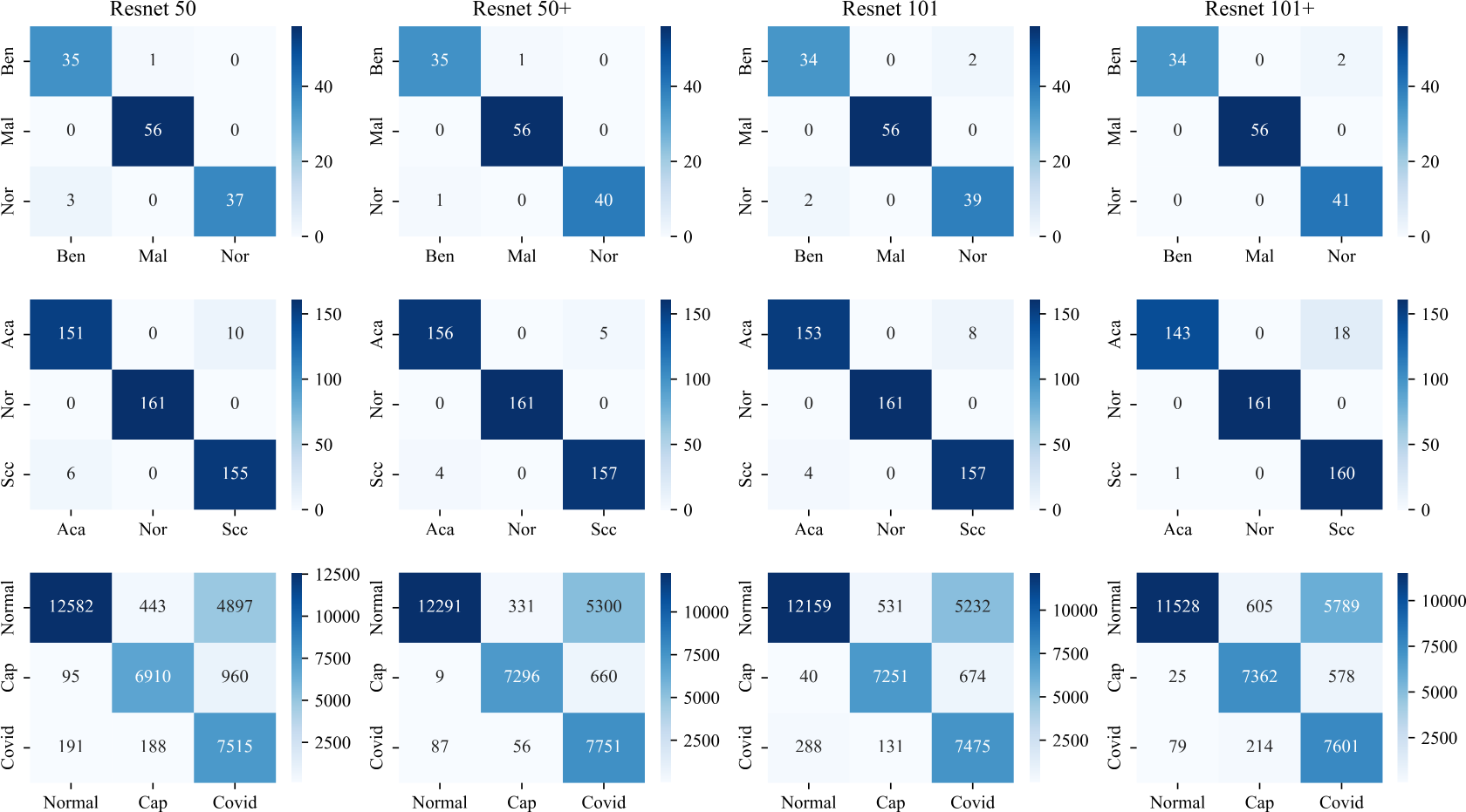}
\caption{{Confusion matrix of the CNN models using IQ-OTCNCCD (\textbf{First row}: Ben, Mal, Nor represent benign, malignant, and normal case, respectively) and LC25000 (\textbf{Second row}: Aca, Nor, and Scc represent lung adenocarcinoma, benign, and small cell lung cancer, respectively) and COVIDxCT (\textbf{Last row}) dataset.}}
\label{fig:cms}
\end{figure*}

\begin{figure*}
\centering
\includegraphics[width=0.98\linewidth]{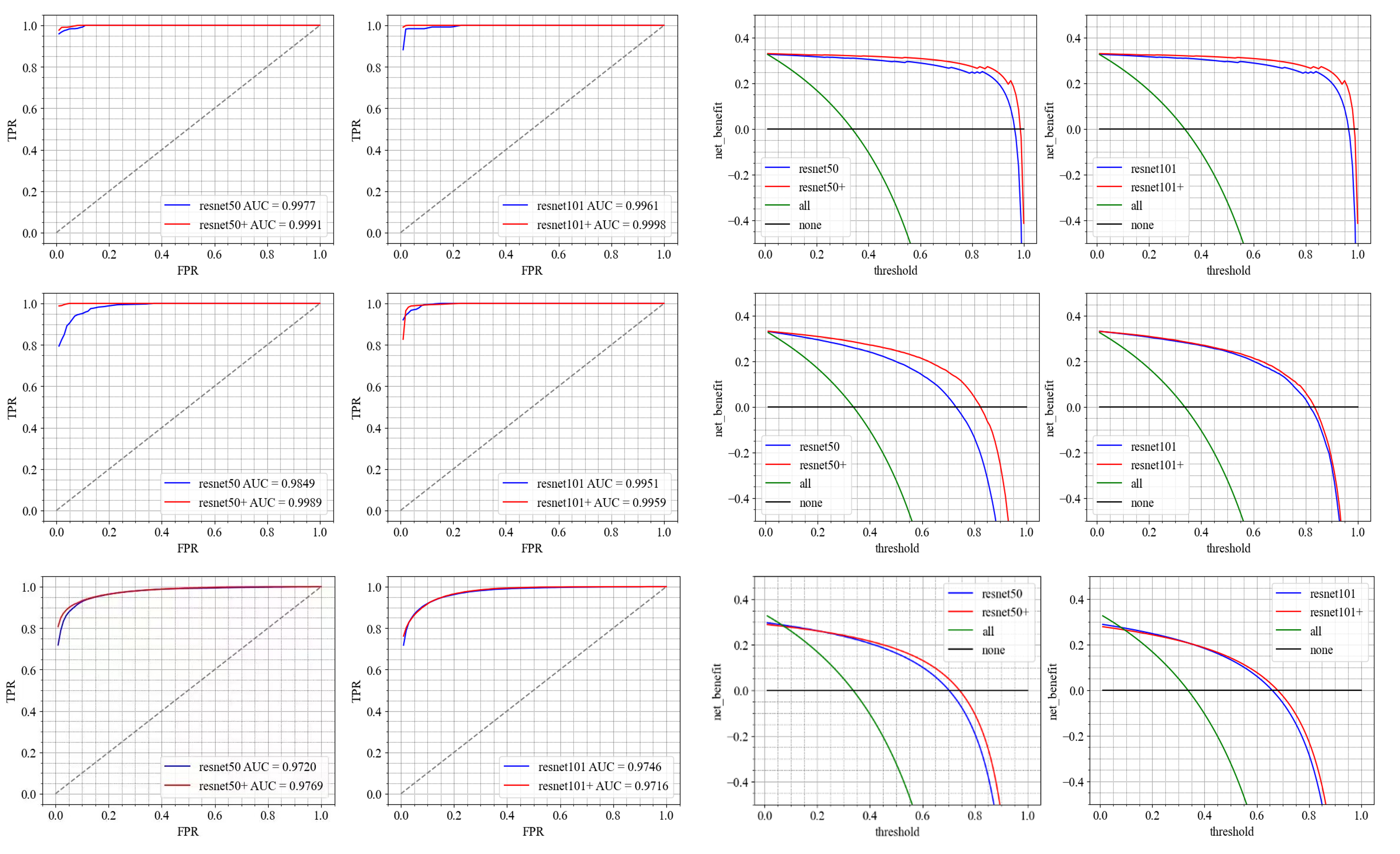}
\caption{Examples of DCA and ROC curves for baselines and our model (\textbf{First row}: IQ-OCTNCCD; \textbf{Second row}: LC25000; \textbf{Last row}: COVIDxCT).}
\label{fig:DCA_ROC}
\end{figure*}

Table \ref{tab:result} reports the results of the classifier model. \textcolor{black}{For lung cancers, the ResNet101+ model performs better compared to ResNet101 in the LC25000 dataset (e.g., 97.52\% ACC vs 96.07\%ACC), while the ResNet50+ model provides higher test metrics compared to ResNet50 in the IQ-OTHNCDD dataset (e.g., 99.25\% ACC vs 96.24\% ACC). Furthermore, both ResNet50+ and ResNet101+ show higher performance metrics compared to original ResNet in LCC (e.g., $\sim 2\%$). For COVID-related dataset, the use of ResNet+ series lead to $\sim 4\%$ improvements in ACC on the ChestXray dataset. In addition, ResNet50+ provides the highest AUC value (0.96), highlighting its potential for X-ray images. In these data sets, the ResNet+ model usually performs better than its baseline (ResNet50 and ResNet101). This finding suggests that the improved architecture with the attention mechanism has a positive impact on the performance of the model.} Figure \ref{fig:cms} shows an example of confusion matrix for classifying the three classes in the LC25000 and IQ-OTHNCDD data set. We observed that ResNet50+ and ResNet101+ outperform their baseline ResNet in predicting normal cases (IQ-OTHNCDD) and small cell lung cancer (LC25000). The AUC values for ResNet50+ and ResNet101+ are not notably different from those of ResNet50 and ResNet101, as illustrated in Figure \ref{fig:DCA_ROC}. However, when evaluated with the DCA curve, ResNet50+ and ResNet101+ show a greater net benefit compared to ResNet50 and ResNet101. In summary, the ResNet50+ and ResNet101+ model demonstrate higher performance compared to baseline ResNet in both lung cancer data sets.

\noindent\textbf{Impact of proposed model in a large dataset}
We evaluated our model on the large COVIDx-CT data set to assess its performance and generalizability over a large dataset. Table \ref{tab:result} presents the experimental results of the model applied to the COVIDx-CT dataset. The ResNet101+ model achieves an accuracy of 81.21\%, demonstrating higher performance compared to other models. Furthermore, Figure \ref{fig:cms} and Figure \ref{fig:DCA_ROC} show the confusion matrix, AUC-ROC and DCA curves for ResNet50, ResNet50+, ResNet101 and ResNet101+. Although the accuracy of the ResNet50+ model is lower than that of its baseline model, the AUC value of ResNet50+ is higher than that of its baseline model. This is due to the imbalance in the dataset classes. Our proposed model can better identify pneumonia and covid, but is not sensitive to non-disease images. However, the DCA curve suggests that ResNet50+ offers less benefit in helping physicians identify pneumonia on medical images compared to ResNet50, due to the greater number of normal images compared to covid and pneumonia images. As illustrated in Figure \ref{fig:cms} and Figure \ref{fig:DCA_ROC}, ResNet101+ demonstrates higher performance metrics (i.e., accuracy and AUC) in classifying normal, pneumonia and covid images compared to ResNet101.

\noindent\textbf{Computational load.} \textcolor{black}{We record the training time (mins) to determine the best validation model for our approach and baselines. As reported in Table \ref{tab:time}, ResNet50+ and ResNet101+ use less computational time to reach their best validation model on the LC25000 and IQ-OCTNCCD datasets, while consuming $\sim 5\%$ more time compared to the original ResNet series on the ChestXray dataset. These findings highlight the impact of CBAM and ResNet-D in predicting lung cancer classes in a short time.} {Furthermore, we provide the inference latency (ms, average $\pm$ standard deviation) for our approach and baselines, as reported in Table \ref{tab:inference_time}. ResNet+ shows a slightly higher inference time (e.g. 12.85ms vs. 11.07ms) for each sample due to its CBAM and ResNet-D architectures.  }

\begin{table}[]\scriptsize
    \centering
    \setlength{\tabcolsep}{5pt}
    \caption{\textcolor{black}{Computational costs (mins) based on a single RTX 2060.}}
    \renewcommand{\arraystretch}{1}
    \begin{tabular}{c|ccc|cc}
    \toprule
       Alg.&LC25000&IQ-OCTNCCD&LCC&ChestXray&COVIDxCT\\
       \midrule
         ResNet50+&103.5 &61.49 &131.48&154.16 &2377.9 \\ 
         ResNet101+&180.09 &80.15 &224.83&268.23 &4066.3 \\
         ResNet50&110.73 &65.2 &132.3&147.39 &2901.3\\ 
         ResNet101&192.67 &93.07 &230.20&248.11 &5048.262\\
         
         \bottomrule
    \end{tabular}
    
    \label{tab:time}
\end{table}

\begin{table}[!ht]\scriptsize
    \centering
    \setlength{\tabcolsep}{3pt}
       \renewcommand{\arraystretch}{1.2}
    \caption{{Inference time (ms, average $\pm$ standard deviation) of ResNet+ and ResNet series. }}
    \begin{tabular}{c|ccc|cc}
    \toprule
       Alg.&LC25000&IQ-OCTNCCD&LCC&ChestXray&COVIDxCT\\
       \midrule
         ResNet50+&11.85±1.54 &7.63±0.92 &15.32±1.84 &17.95±2.15 &14.41±1.13 \\ 
         ResNet101+&13.06±2.53 &9.82±1.18 &16.13±3.02 &19.14±3.62 &15.35±1.19 \\
         ResNet50&11.07±1.33 &6.52±0.78 &13.23±1.59 &15.74±1.77 &12.13±1.82\\
         ResNet101&12.27±2.31 &9.31±1.12 &15.02±2.76 &17.81±2.98 &14.83±1.58\\
         \bottomrule
    \end{tabular}
    \label{tab:inference_time}
\end{table}



\noindent\textbf{Ablations on CBAM and Resnet-D} \textcolor{black}{We further validate the usefulness of our CBAM module using the five data sets. As reported in Table \ref{tab:ablation}, the use of both CBAM and ResNet-D produces the highest ACC compared to the use of CBAM or ResNet-D alone (e.g., 99.44\% vs 96.82\% vs 97.82\% in the LCC dataset). These results highlight the potential of the CBAM and ResNet-D modules.}

\begin{table}[]\scriptsize
    \centering
    \caption{\textcolor{black}{Accuracy (\%) according to presence \ding{52} or absent \ding{56} of CBAM and ResNet-D module (ResNet101).}}
    \renewcommand{\arraystretch}{1}
    \setlength{\tabcolsep}{3.8pt}
    \begin{tabular}{cc|ccc|cc}
    \toprule
       CBAM&ResNet-D&LC25000&IQ-OCTNCCD&LCC&ChestXray&COVIDxCT\\
       \midrule
        \ding{52}&\ding{56}&96.82 &98.12 &96.82&81.89 &80.29 \\
         \ding{56}&\ding{52}&96.94 &98.34 &97.82&82.37 &80.59 \\
         \ding{52}&\ding{52}&97.52 &98.50 &99.44&83.01 &81.21 \\
         \bottomrule
    \end{tabular}
    
    \label{tab:ablation}
\end{table}

\noindent\textbf{Convolution architectures and shortcuts.} {We further validate the impact of  swapping convolution orders (SCO), replacing 7×7 convolutions, and modifying shortcuts. As reported in Table \ref{tab:ablation_2} (with ResNet50 as backbone), without modifying shortcuts leads to $\sim2\%$ performance decrease on LCC, LC25000, IQ-OCTNCCD datasets, while using SCO and replacing 7x7 convolutions can increase the performance on LCC, LC25000, IQ-OCTNCCD and ChestXray datasets. With all components, ResNet50+ provides the best overall performance (e.g., ACC: 99.99\% on LCC dataset). }

\noindent\textcolor{black}{\textbf{Multi-modality data.} {We validate our model in a multi-modal setting (e.g., CT with pathology images). The LC25000 and IQ-OTHNCDD datasets were combined to train and validate our model. Table \ref{tab:result2} reports the performance metrics. ResNet50+ and ResNet101+ provide higher test metrics such as ACC (e.g., 96.63\% vs. 94.18\% with ResNet50+ and ResNet50), F1 (93.69\% vs. 92.83\% with ResNet50+ and ResNet50), highlighting their potential for multimodality image classification.}}

\begin{table} \scriptsize
\caption{\textcolor{black}{Performance metrics (\%) using the multimodality datasets}}
\renewcommand{\arraystretch}{1}
 \setlength{\tabcolsep}{11.2pt}
\begin{tabular}{c|ccccc}
\hline
Model& ACC & PRE & REC & F1 & AUC\\
\hline
ResNet50+&96.63&93.78&93.62&93.69&0.978  \\
   ResNet101+    &93.76&92.15&88.13&88.83&0.945   \\
ResNet50     & 94.18&92.82&92.22&92.83&0.961   \\
ResNet101    & 94.43&93.26&86.63&86.97&0.933   \\
    \hline
                            
\end{tabular}
\label{tab:result2}\vspace{0.5cm}
\end{table}

\begin{table}[]\scriptsize
    \caption{
    {Accuracy (\%) according to presence \ding{52} or absent \ding{56} of swapping convolution orders (SCO), replacing 7x7 convolutions (RC) and modifying shortcuts (MS).}}
    \setlength{\tabcolsep}{4pt}
    \renewcommand{\arraystretch}{0.8}
    \begin{tabular}{ccc|ccc|cc}
    \toprule
       SCO&RC&MS&LCC&LC25000 &IQ-OCTNCCD &ChestXray & COVIDxCT \\
       \midrule
        \ding{52}&\ding{52}&\ding{52}&99.99&98.14 &99.25 &87.98&78.45  \\
         \ding{56}&\ding{52}&\ding{52}&99.10&97.20 &98.64 &84.13&77.0   \\
         \ding{52}&\ding{56}&\ding{52}&99.19&97.88 &98.60 &85.25&77.5   \\
         \ding{52}&\ding{52}&\ding{56}&97.82&96.80 &97.96 &87.01&78.01   \\
         \bottomrule
    \end{tabular}
        \label{tab:ablation_2}
\end{table}

\noindent\textbf{CBAM in skin cancer classification.} {In \cite{su2025relation}, they argued that CBAM is useful for skin cancer classification tasks. Thus, we validate the usefulness of CBAM in our ResNet50+ using the ISIC2018 \cite{tschandl2018ham10000,codella2019skin} dataset. Specifically, ISIC2018 has seven classes, with a train set of 10015 samples, and a test set of 1512 samples. We use the same implementation settings as described in Section \ref{SS:implementation} for experiments. As reported in Table \ref{tab:ISIC2018}, ResNet50+ provides a higher PRE and F1 score (49.42\%, 38.65\%) compared to ResNet50 (47.53\%, 38.33\%). In addition, ResNet50+ yields a higher AUC value of 0.697 than ResNet50 (0.678). These results highlight the usefulness of CBAM in similar tasks.}

\begin{table} \scriptsize
\caption{{Summary of the performance metrics (\%) for classifying the test images of ISIC2018 dataset.}}
\renewcommand{\arraystretch}{1.2}
 \setlength{\tabcolsep}{11.2pt}
\begin{tabular}{c|ccccc}
\hline
Model& ACC & PRE & REC & F1 & AUC\\
\hline
ResNet50+&67.98&49.42&37.12&38.65&0.697  \\
   ResNet101+    &70.76&50.15&43.13&44.83&0.703   \\
ResNet50     & 68.12&47.53&36.19&38.33&0.689   \\
ResNet101    & 69.37&45.9&42.77&43.69&0.693   \\
    \hline
     
\end{tabular}
\label{tab:ISIC2018}
\end{table}

\noindent\textbf{Attention based ResNet and hybrid approaches.} {Table \ref{tab:comparison} reports the test metrics for ResNet+ and recent attention-based ResNet \cite{liu2022crorelu} and hybrid model \cite{talib2024transformer}. In \cite{liu2022crorelu}, they proposed a novel ResNet based model equipped with channel
attention and cross-space activation module. In \cite{talib2024transformer}, they combined transformer with cnn networks to enrich the image features used for lung cancer prediction. For example, ResNet50+ provides higher test metrics (e.g., 98.14\% F1) compared to other baselines. These results highlight the potential of CBAM with ResNet-D.}

\begin{table} \scriptsize
\renewcommand{\arraystretch}{1.4}
\caption{{Performance metrics (\%) on LC25000 dataset.}}
 \setlength{\tabcolsep}{7pt}
\begin{tabular}{cc|ccccc}
\hline
&Model& ACC & PRE & REC & F1 & AUC\\
\hline
\multirow{4}{*}&ResNet50+& 98.14   & 98.14    & 98.14 & 98.14&0.998  \\
   &ResNet101+  & 97.52   & 97.53    & 97.52 & 97.52&0.996   \\
&CNN+Transformer \cite{talib2024transformer}     & 98.39&98.00&97.00&98.00&-   \\
& SENet50 \cite{liu2022crorelu}&99.96&99.87&-&-&- \\
\hline                     
\end{tabular}
\label{tab:comparison}
\end{table}

\begin{figure}
    \centering
\includegraphics[width=0.95\linewidth]{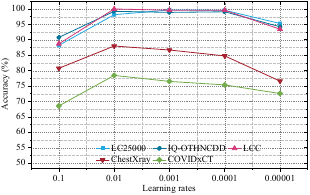}
    \caption{{Accuracy (\%) of test samples  derived from five datasets with learning rates.}}
    \label{fig:learning_rate}
\end{figure}

\noindent\textbf{Learning rates.} {We validate the impact of learning rates using five datasets. As illustrated in Figure \ref{fig:learning_rate}, a higher learning rate (i.e., 0.1) or a lower learning rate (i.e., 0.00001) degrades the performance $>5\%$. Optimal starting point can be set to [0.01, 0.001].}

\section{Discussion and Conclusion} \label{conc}
The popularization of early detection of lung cancer is a challenge, which is limited by many factors, including medical costs and detection speed. DL models have their own advantages in this field, but their performance is limited by the imbalance of dataset classes and poor model generalization. Therefore, in the framework of the lung cancer classification model proposed in this study, CBAM is added to enhance the generalizability of the model. Furthermore, data augmentation is used to solve the problem of class imbalance in the data set, and the ResNet-D module is used to reduce model operation cost and feature loss. Our proposed ResNet50 + model achieved an accuracy of 99. 25\% in the IQ-OTH / NCCD dataset. This result is consistent with previous studies that achieved an accuracy range of 98.32-99.10\% \cite{discussion1,discussion2}. Compared with similar studies, our proposed model has advantages in classification performance. This shows that the classification model we proposed has certain clinical practical value. However, there are still some limitations to our study. First, our proposed model was only validated in the lung task. More data sets are necessary to verify the effectiveness of the model. Second, when we scaled the model to a large dataset, our proposed ResNet50+ model performed worse than its standard model, which may be due to the gap caused by the unbalanced distribution of data set samples. \textcolor{black}{Moreover, in clinical applications, privacy concerns pose a challenge in collecting large amounts of training data, potentially limiting our model use. Federated learning offers a solution by allowing privacy-preserving data sharing \cite{chaddad2024federated}. Additionally, incorporating interpretable AI can enhance the transparency of our model, thereby increasing clinician confidence in its predictions \cite{chaddad2024generalizable}.}

As conclusion, we developed a ResNet+ to predict lung cancer images. The proposed model demonstrated higher performance metrics compared to baseline models using public datasets. The diversity and representativeness of the data set can affect the generalizability of the model. Future work will expand the sources and types of data to improve the adaptability of the model to different clinical tasks. In addition, we plan to explore multimodal approaches that combine clinical data and biomarkers to further improve lung cancer classification performance metrics.

\section*{Acknowledgments}
All authors declare that they have no known conflicts of interest in terms of competing financial interests or personal relationships that could have an influence or are relevant to the work reported in this publication.

\bibliographystyle{unsrt}
\bibliography{reference}

\end{document}